# Emerging Devices and Packaging Strategies for Electronic-Photonic AI Accelerators

Nicola Peserico[1], Thomas Ferreira De Lima[2,3], Paul R. Prucnal[2], Volker J. Sorger[1,*]

[1] *Department of Electrical and Computer Engineering, George Washington University, Washington, DC 20052, USA*
[2]*Currently with NEC Laboratories America Inc, Princeton NJ 08540, USA*
[3]*Department of Electrical and Computer Engineering, Princeton University, Princeton NJ 08544, USA*
**Email: sorger@gwu.edu*

**Abstract:** The field of mimicking the structure of the brain on a chip is experience much interest driven by the demand for machine intelligent applications. However, the power consumption and available performance of machine-learning accelerating hardware still leaves much desire for improvement. Application specific integrated circuits (ASIC) including emerging devices offer intriguing algorithm-hardware homomorphism. In this letter, we share viewpoints, challenges, and prospects of electronic-photonic neural network accelerators. Combining electronics with photonics offers synergistic co-design strategies for high-performance AI ASICs and systems. Harnessing signal processing advantages from photonics such as processing parallelism deployed in tensor operation processors, high signaling speed, high fan-out enabling neuron interconnectivity in neuromorphic architectures, and combining them with electronic logic control and data storage is an emerging prospect. However, the component library leaves much to be desired for and is challenged by enormous size of photonic devices or challenge for retention-of-state. Within this context, we will review the emerging electro-optic materials, functional devices, and systems packaging strategies that, when realized, provide significant performance gains and fueling the ongoing AI revolution. We discuss challenges that the field faces and offer a solution roadmap utilizing heterogeneous technology integration leading to a stand-alone photonics-inside AI ASIC 'black-box' for streamlined plug-and-play board integration in future AI processors.

## 1. Introduction

Neuromorphic Photonics (NP) has gathered increasing interest in recent years owing to its potential to disrupt performance of classes of applications that conventional digital processors are challenged with attain [1]. Namely, artificial intelligence-based applications can be mapped to either tensor operation ASICs [2] or brain-inspired circuits such as neural networks, programmed via a learning procedure. Photonic circuits are well suited to high-performance signal processing implementations of neural networks for two predominant reasons: interconnectivity and high-bandwidth of linear and nonlinear operations. Neural networks require a large web of independent connections between layers containing individual neurons. Simply put, each connection between a pair of neuron needs a scalar weight value (or synaptic weight). If the output of a neuron layer can be represented by a vector, each neuron in a subsequent layer masks that vector by applying a dot product with a weight vector. This results in a layout of interconnections that can be represented as a matrix-vector multiplication, or, for fully two-dimensional data such as for image processing, as a matrix-matrix multiplication.

It is possible to implement this multiplication via passive interferometric devices with tunable elements. For example, if an input vector is represented by wavelength-division multiplexed (WDM) optical signals, a weight can be applied by a sequence of tunable microring resonator weights, called a weight bank [3]. In a scheme called broadcast and weight [4], these resonators determine the interconnectivity as well as the weights. The nonlinearity can also be achieved with an electrically tunable interferometer such as a microring resonators (MRR) or a Mach Zehnder interferometers (MZI) [5].

Silicon photonics (SiPh) offers to integrate a high density of optoelectronic devices combined with high quality passive components. It leverages the decades of research and development from CMOS fabrication lines [6], and is gaining momentum as a dominant integrated photonic platform driven by cost and chip performance alike [7]. The main limitations of a SiPh platform for inference and machine-learning architectures lay on the lack of complex on-chip electronic circuitry for calibration and control, as well as solutions for generating light on-chip efficiently. In order to progress from early system demonstrations to fully integrated processors, NP systems require new materials and technologies. For example, correcting fabrication variability in a post-fabrication step can reduce heat dissipation and the amount of current needed for tuning. Memory circuits that are able to interact directly with light can enable more agile reconfiguration in the processor, as well as self-learning capabilities. Finally, information need to be transferred through digital interfaces between electronic and photonic processors with minimal heat dissipation, which can be achieved with integration of optical sources and high-efficiency modulators. In this paper, we discuss potential solutions for these limitations, outlining key emerging devices that can have an extraordinary impact on the performance of neuromorphic photonic processors.

## 2. Moving Beyond Silicon Photonics

In an electronic-photonic AI accelerator, the function and performance of a neural network a tensor core processor rely strongly on the capability of realizing and implementing (i.e. WRITE, READ, RESET, STORE operations) the weights and biases. These have to be applied to each optoelectronic (or electronic) device representing either the weights during the machine learning training step, or the trained weights during inference tasks. Depending on the specific application, this requires precise phase and/or amplitude tuning (depending on the weighting scheme) of each waveguide segments to realize the bit resolution of the selected machine learning task.

We can distinguish between weight requirements and hence implementation options for classification versus machine learning [8] ('training', i.e. such as in cloud services e.g. Amazon Web Service) applications of the underlying electronic-photonic ASIC accelerator (fig.1a); while the update rate for classification is seldom (i.e. whenever a new data set is available producing updated weights) during training step, the weights and biases need to be updated constantly. This naturally leads to non-volatile state-retention for classification as compared to rapid (faster-the-better) WRITE-RESET-REWRITE etc cycles for cloud training applications. Specifically at the network edge the size-weight-area-performance (SWAP) requirements of AI systems are demanding, including high energy efficiency, hence stressing the non-volatile capability of the deployed weight-memory; that is, once the weights are WRITTEN, they (ideally) are zero-power consuming static functions. For memory options, electronics offers a variety of memory options with the trade-off between WRITE-speed and energy vs. READ latency and include cache, SRAM, DRAM, FLASH (in increasing retention-time order). Interestingly, a brief analysis suggests a 100x superior potential of photonic memory over state-of-art SRAM with respect to data baud rate (speed) and memory access energy; in brief, an SRAM has an access latency of 0.3ns costing about 100fJ/access. A photonic memory based on phase-chance-materials (PCM), once WRITTEN, requires only the photon creation and detection energies. The minimum power of foundry-based PIC detectors in the C-band, for example, are about 50nW for signals above 30GHz. Assuming a 1% efficiency for the laser wall-plug efficiency and optical losses on the PIC and coupling to the PIC [1-2dB per coupler], a memory READ (access) energy of a PCM-written photonic random-access memory (P-RAM) takes <1fJ/access for an on-off-keyed (OOK) signal at 30GHz data rates, or, about 10fJ/access for a higher bit resolution (e.g. PAM16 for a 4-bit ML classifier). Thus, a generic photonic link offers MAC operations and memory access of 10-100x higher MAC/s/J/access than SRAM. Following this potential for PIC-based MAC acceleration, electro-optic reconfigurable photonic integrated circuits (PIC) have been predicted [9] and demonstrated [5] to process the repeating convolution-underlying MAC operation (multiply-accumulate) in inference tasks on off-chip trained kernels.

Exploring the memory-(active)material relationship further; PCMs have recently shown promising capability, both as amplitude and phase modulation [10], exploiting the non-volatile switching between their amorphous and crystalline states. In conventional SiPh, this is achieved by either heating elements placed near a waveguide, or by doping the waveguide with a p-n junction. These approaches rely on the thermo-optic effect and the free-carrier depletion (FCD) effect, respectively. But, since they require constant active electrical current, their energy-efficiency is limited. PCMs are non-volatile and requiring no active energy consumption, making them ideal for fabrication variation trimming as well as optical memory for edge AI applications. The FCD effect in silicon being weak requires high voltages to achieve significant index modulation. This inefficiency is problematic because neurons and tensor weight updates require a sufficient modulation so that small signals can span all features of the nonlinear activation function [11]. Better results, that is micrometer compact MZI modulators at GHz speeds could be obtained using free-carrier-based electro-optic modulation such as in the material Indium Tin Oxide (ITO) [12], as enabled by unity-strong index modulation [13].

### *2.1. Emerging Materials for Non-volatile Optical Memory*

**Why is memory important?** Memory is key element of all the modern computer, as CPU keeps transferring data and instructions from and to the memory side of the computer. On the optical side, non-volatile optical materials have the potential to play the same role, allowing to store phase and amplitude variation on a photonic circuit. As NP links a large web of interconnect synapses, having a precise and non-volatile control of the synaptic weights is essential. The PCM recently became one of the most popular and promising active materials for the realization of this type of non-volatile multilevel random access photonic memory (P-RAM).

In recent days, the widely studied PCMs include transition metal oxide, chalcogen-based and antimony based PCMs, such as vanadium $VO_2$, $Ge_2Se_2Te_2$ (GST), $Ge_2Sb_2Se_4Te_1$ (GSST), $Ge_2Sb_2Se_2$ (GSSe), and $Sb_2Se_3$. All those materials are embedded into silicon-based photonic devices whose phase can be reversibly changed between crystalline and amorphous via appropriate heating processes. The photonic properties of the material between the two phases are significantly different with distinct refractive index (n/k) contrast which divides nonvolatile programmable PIC memories into two groups: phase-shifting modulation, and amplitude modulation.

For the phase-only modulation, a variety of PCMs is embedded into resonate-based PICs such as MZI, micro-ring resonators, and directional couplers [14, 15, 17]. Vanishingly small insertion loss, large index contrast (□n) PCMs such as GSST4 and $Sb_2Se_5$ are preferred in the telecommunication bands that confers low insertion loss, small footprint, and non-static power consumption.

For amplitude-only modulation, the variety of large index contrast (□k) and small insertion loss in amorphous state PCMs are covered over the silicon-based waveguide such as GSSe and GST [10, 16].

To trigger the structural transition of PCM, the local annealing is required to apply on the material which normally relies on the external laser heating, or the micro-resistive heater to actuate structural phase transition via multi-electrical pulse combinations. From the perspective of overall footprint and phase transition energy efficiency, external laser heating is a better option compared to micro-heating since the pump laser can be directly guided to the target material as a heating source without extra structure on-chip needed. On the other hand, for micro-resistive heaters such as Tungsten, Graphene, or ITO heaters, though they need extra space for contact pads and routing, the programming setup is relatively simpler compared to the external heating. This is especially true for the large-scale NPs which require a large amount of PCM memories for the weight banks, as they could be easily controlled through electrical control unit instead of multiple modulated external laser sources. A mix of the two approaches might combine the best parts of both, having larger post-fab trimming with laser heating, and fast tuning with local electrical heaters. All these developments of PCM-based PRAMs are the fundamental steps towards the fast, lower-power consumption, high bit resolution on-chip photonics memory for neuromorphic computing.

| | |
|---|---|
| Non-Volatile Edge | Volatile (10GHz) Cloud |
| High | Medium |

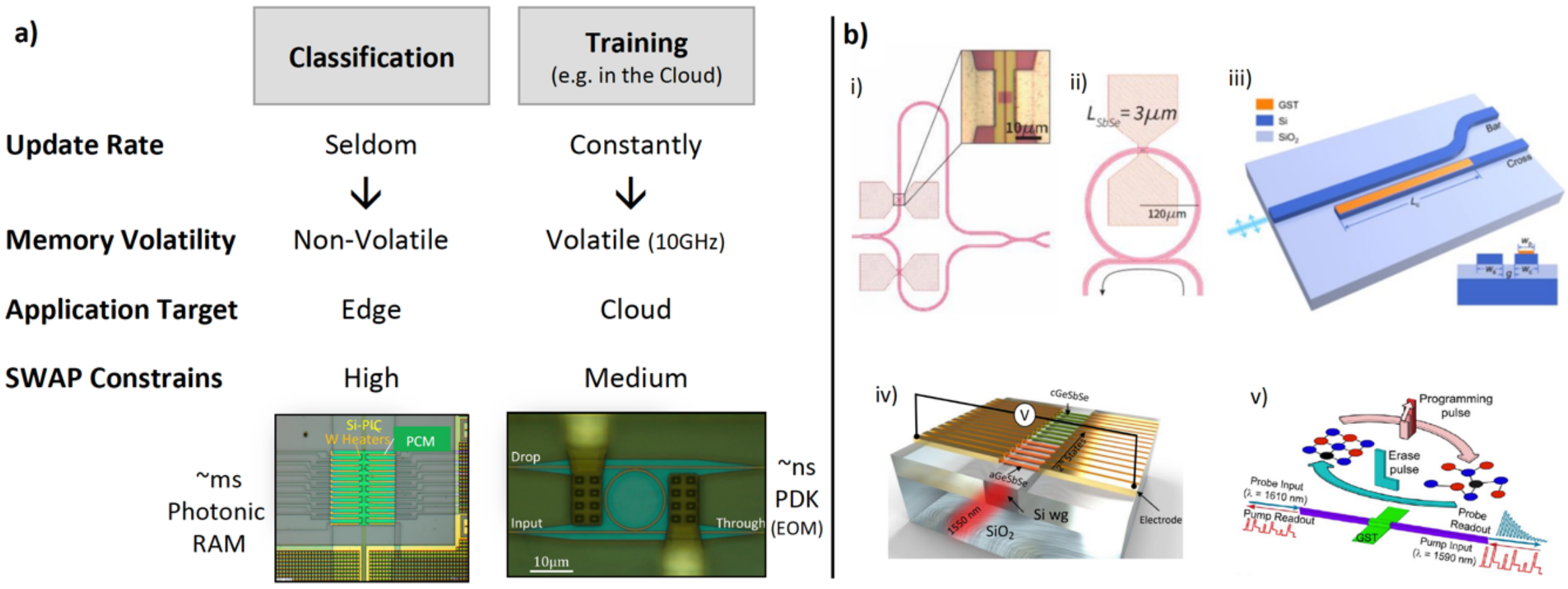

**Figure. 1.** Performance of the AI ASIC weights and biases rely on the material of the reconfigurable memory devices. (a) Memory requirements depend on the machine learning application such as classification vs. machine-learning training [8]. (b) Photonic random-access-memory (P-RAM) options from the recent literature include multi-state memories [10]. i) An unbalanced MZI with $Sb_2Se_3$ cell [14]. ii) Micro-ring resonator with a $Sb_2Se_3$ [14]. iii) A 1x2 directional coupler with GST cell [15]. iv) Waveguide with GSSe cell and multiple double-sided tungsten heaters [10]. v) Schematic of the laser pulse to amorphized and crystallize the integrated phase-change photonic memory cell [16]. Interestingly, one finds a 10-100 fold higher speed-energy product performance (for the memory READ step) of P-RAMs when compared to SRAM technology.

### *2.2. Efficient Modulator Materials for Silicon Photonics*

Current p-n junction-based SiPh platforms do not support highly-interconnected photonic neural networks unless they use (a) more sensitive modulators, (b) active transimpedance amplifiers (TIAs), or (c) operate at a sub-GHz bandwidth [11]. This occurs because modulators need a large voltage swing to reach the nonlinear threshold in their nonlinear transfer function, which suppresses noise directly between one neural layer and the next – a requirement for cascadable analog links. This swing can either be achieved by increasing optical pump power at the modulator or by providing electric transimpedance gain. However, optical gain is limited by optical nonlinearities in waveguides (and potentially power budgets), whereas transimpedance gain is inversely proportional to the bandwidth of the circuit.

In order to construct O/E/O neurons compatible with SiPh that can operate at >10 GHz, we need a modulator that has simultaneously a low-capacitance (RC limited) and a low-Vpi (e.g. 50 mV). Comparing the $V_\pi L$ parameter, while p-n junctions are limited to 360 V$\mu$m, other heterogeneously devices could reach the wanted performances, such as ITO-based modulator that can reach 95V$\mu$m [13], and ITO-graphene device, that can implement even high-bandwidth modulation, up to more than 130 GHz [18]. Exploiting the properties of these materials allows to create more specific devices such as the Non-Linear Activation Functions for NP.

## 3. System Integration Strategies for Electronic-Photonic AI ASICs

While PCMs help enable an efficient electrical reconfiguration medium, they require precise analog inputs and an electronic control framework. Unless the required analog signals can be generated and detected on-chip from a smaller set of inputs (through e.g. monolithic integration), each component requires its own electrical bondpads for interfacing with an external control chip.

For large-scale neuromorphic circuits, this means the chip's footprint is dominated by electrical routing and thermal constraints. Because photonic devices are limited in size by the lightwave's wavelength, PICs are fabricated using lithographic methods from previous node generations as current CMOS. As a result, PICs are expected to have a larger area than CMOS ASICs, and can also serve as an interposer between the CMOS ASIC and the PCB.

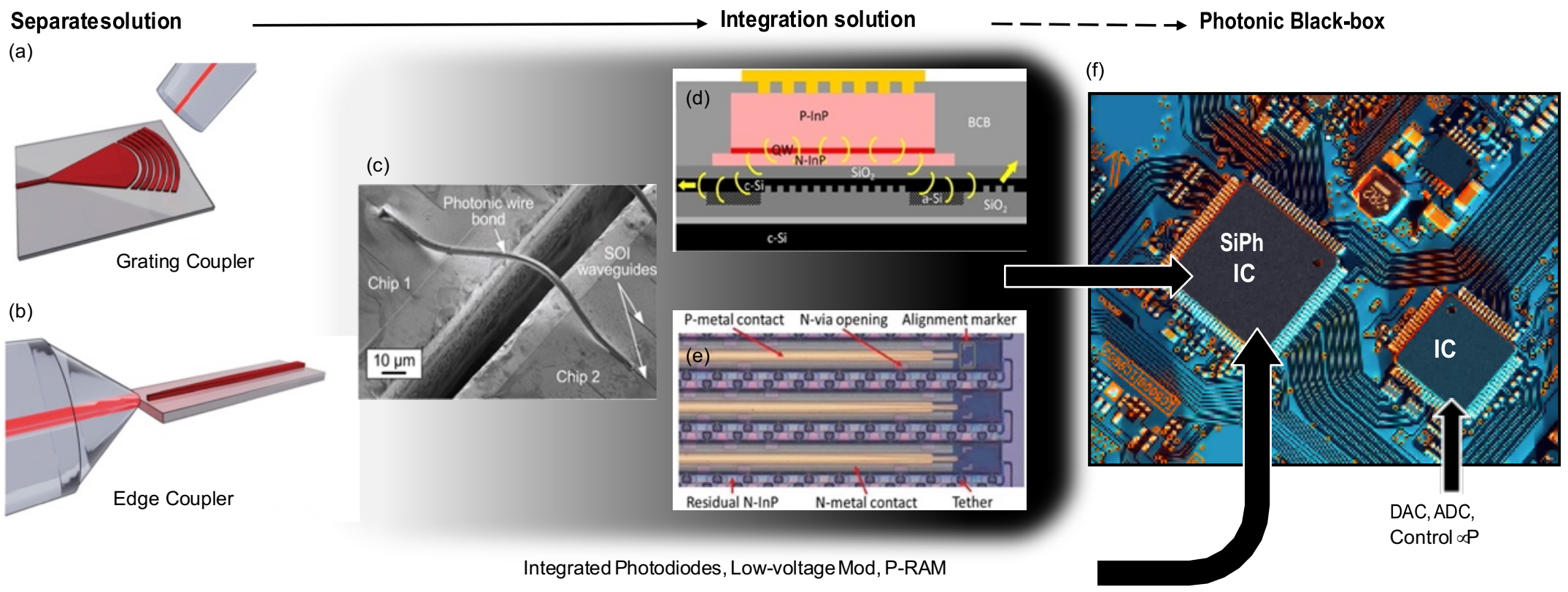

**Figure. 2.** Example of I/O integration for PIC. a-b) Well-known coupling structures, such as Grating Couplers and Edge couplers [19]. c) Example of Photonic Wire Bonding connecting laser to PIC [20]. d) Back-side-on-BOX heterogeneously integrated III-V- on-silicon [21]. e) Hybrid integrated semiconductor using Micro-Transfer-Printing [22]. f) Graphical representation of the Photonic Black-box, where all the optical components are inside a single IC having just electrical I/O.

While electrical connectivity can borrow industry standards from the semiconductor manufacturing industry, optical interfaces are less matured, yet recent development give promise for an efficient and salable optical I/O driving ultimate PIC-electronic integration, as shown in fig. 2. Today, signal I/O is generally addressed by utilizing coupling structures such as grating couplers, or edge couplers. These solutions show low insertion loss (as low as 0.5dB [19]) but still rely on sub-micron alignment of fibers. More recently, promising research on optical source integration surfaced. One solution is termed photonic wirebonding [20], which can connect arbitrarily placed devices on the same interposer or substrate. This approach has the most potential in the short term, enabling integration of existing solutions into one single platform.

Full integration of lasers on silicon has been demonstrated by a different research groups. Notable examples include Back-side-on-BOX heterogeneously integrated III-V-on-silicon [21], quantum dots on silicon [23], and hybrid integrated semiconductor [22]. Integrated optical sources on silicon represents a substantive push for AI ASIC hardware and subsequent applications, because it can be miniaturized and deployed in the field, where edge AI processing is the most bottlenecked, while supporting more stringent vibration and temperature fluctuation requirements. Moreover, chips can be assembled at-scale (volume) without requiring advanced optical packaging, which will likely offset the increased fabrication cost due to heterogeneous integration [24]. Our vision and ongoing explorations are to co-integrate photonics-inside, fully-packaged 'black-box' photonic ASIC accelerators on the same printed circuit boards (PCB) as electronic ICs (fig. 2(f)). This will revolutionize not only photonic AI hardware prototyping, but also ripple through the entire PIC community; offering an stand-along photonic system (including source, programmable circuit, and O-E back-end) results in an 'photonic-hidden' module that electronic circuit designers can use as a plug-and-play design module without having to have much (possibly ’any’) understanding of the optical details.

## 4. Concluding Remarks

Here we review latest advances in photonic programmable circuits and devices for realizations of electronic-photonic ASICs for machine-learning (ML) applications. Here, we discuss relevant material- and device design options, which are the underlying fabric of these emerging mixed-signal AI and ML processors processor. We analyze the requirements for the machine-learning weights and biases by distinguishing between classification (inference) vs. training applications. For each, we find a different set of device and hence ’active’ (programmable) performance requirements. For network edge applications with rare weight-updating, implementing non-volatile photonic random-access-memory (P-RAM) suggests a 10-100 times higher baudrate-energy performance for WRITTEN weights, i.e. the memory READ step. Furthermore, we show examples of how the nonlinear activation function can be efficiently realized for neuromorphic ASICs, for which micrometer compact ITO-based modulators show great promise being 10,000 more compact than modulators based on thin-film Lithium Niobate. Finally, we share our vision and ongoing effort of developing the first photonics-inside (and hidden from the electronic circuit designer) fully packaged photonic AI processor. This co-design strategy leverages recent developments in photonic-wirebonding to same-chip integrate lasers onto silicon PICs. With this, we believe, the future for photonic-electronic ASICs is rather ’bright’ as we are just starting to explore fully co-packaged AI systems. However, CMOS foundries still do not allow materials such as most PCMs and ITO, due to low demand and process compatibility issues. Future work should face these challenges while considering new materials. Moreover, integration of lasers source still requires additional steps in the fabrication process, still facing mass production limitation. As it stands, solutions for neural network weights or kernels for tensor operation accelerators (including convolution acceleration) are being realizable in the prototyping stages at-present. Indeed, some of these challenges can be overcome by the increasing demand in AI hardware and circuits, which should drive implementations of exploratory pilot lines in the foundries that offer material-device-packaging opportunities beyond silicon-only circuits.